\documentclass[
 aip,
 jmp,
 amsmath,amssymb,
 reprint,
]{revtex4-1}

\usepackage{graphicx}
\usepackage{dcolumn}
\usepackage{bm}

\begin{document}
\preprint{}
\title{Empirical Analysis of the Online Rating Systems }
\author{Xin-Yi Lu}
\affiliation{Research Center of Complex Systems Science, University of Shanghai for Science and Technology, Shanghai 200093, P. R. China}
\author{Jian-Hong Lin}
\affiliation{Research Center of Complex Systems Science, University of Shanghai for Science and Technology, Shanghai 200093, P. R. China}
\author{Qiang Guo}
\affiliation{Research Center of Complex Systems Science, University of Shanghai for Science and Technology, Shanghai 200093, P. R. China}
\author{Jian-Guo Liu}
\email[]{liujg004@ustc.edu.cn}
\affiliation{Research Center of Complex Systems Science, University of Shanghai for Science and Technology, Shanghai 200093, P. R. China}

\date{\today}

\begin{abstract}
This paper is to analyze the properties of evolving bipartite networks from four aspects, the growth of networks, the degree distribution, the popularity of objects and the diversity of user behaviours, leading a deep understanding on the empirical data. By empirical studies of data from the online bookstore Amazon and a question and answer site Stack Overflow, which are both rating bipartite networks, we could reveal the rules for the evolution of bipartite networks. These rules have significant meanings in practice for maintaining the operation of real systems and preparing for their future development. 
We find that the degree distribution of users follows a power law with an exponential cutoff. Also, according to the evolution of popularity for objects, we find that the large-degree objects tend to receive more new ratings than expected depending on their current degrees while the small-degree objects receive less ratings in terms of their degrees. Moreover, the user behaviours show such a trend that the larger degree the users have, the stronger purposes are with their behaviours except the initial periods when users choose a diversity of products to learn about what they want. Finally, we conclude with a discussion on how the bipartite network evolves, which provides guideline for meeting challenges brought by the growth of network.
\end{abstract}
\maketitle
\section{Introduction}
In past years, evolving networks have arose the interests of lots of researchers, who study the evolving networks in diverse fields,\cite{temporal,Scale} such as person-to-person communication,\cite{email1,email2} one-to many information dissemination,\cite{information1,information2} neural and brain networks,\cite{brain1,brain2} and ecological networks.\cite{eco1,eco2} Derek Price first studied the growth of citation networks for scientific papers.\cite{Price} He found that the more citations a paper received, the more chances it would be cited in the future, which was called "cumulative advantage" by him.\cite{Price} Barab¡äasi and Albert later studied the World Wide Web, where they found a power-law degree distribution\cite{Barabasi1,Barabasi2} and proposed the well-known model of network growth, named as Preferential attachment(PA). A great number of other researchers such as Strogatz\cite{Strogatz}, Huang\cite{Huang}, Liu\cite{liu2}, D¨ªaz\cite{chaos} also do empirical studies to learn about dynamics of complex networks, building solid foundation for analyzing the development of bipartite networks.

\begin{figure}[h]
\center\scalebox{0.65}[0.65]{\includegraphics{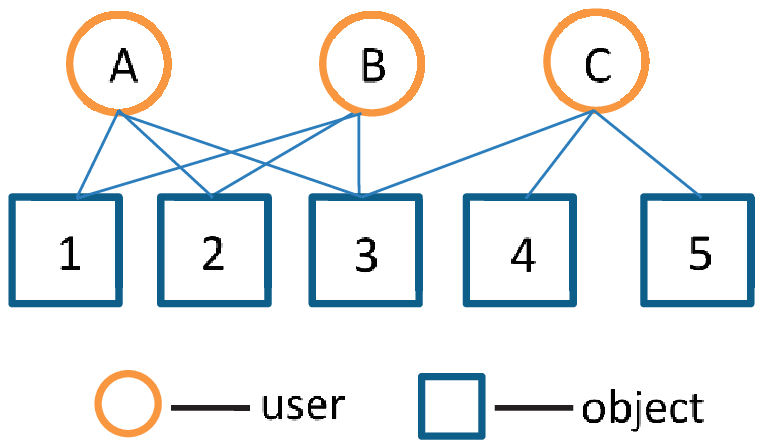}}
\caption{(Color online) A bipartite network with nodes in two types: U=\{A, B, C\}and V=\{1, 2, 3, 4, 5\}. Each edge represents the rating that the object acquired from the user.}\label{net22}
\end{figure}

Interaction between a large number of entities are common in natural and man-made systems. In order to study these interaction, real systems are mathematically represented as graphs, consisting of pairs of nodes and a set of edges, where nodes represent the entities and the edges represent the interaction between entities.\cite{CRC} However, the previous studies mainly focus on networks with nodes of same type and pay little attention to the evolving networks with other structure, like bipartite networks. Bipartite networks conclude nodes in two types: $U=\{u_{1},u_{2},...,u_{n}\}$ and $V=\{v_{1},v_{2},...,v_{m}\}$ where edges only exist between nodes of different types,\cite{CRC} as shown in Figure~\ref{net22}. The most common evolving bipartite networks are e-business networks, which update frequently and combine with the human dynamics. Thus, in this paper, we study the evolving rating networks deeply from several aspects, such as the growth of networks, the user degree distribution,  the popularity of objects and the diversity of user behaviours, by examining the empirical data of Amazon and Stack Overflow.
Medo once used a model based on Preference Attachment to analyze the citation network.\cite{Medo} According to his study, we apply the model in investigating the evolution of popularity for objects and use the information entropy to study the user behaviours. Studying properties of such networks can help us better understand the evolving direction of bipartite networks and capture the features of user behaviours.

The rest of this paper is organized as follows. In Sec. II, we analyze statistical properties of data sets of Amazon book ratings and the Stack Overflow favourite posts to display the basic situation of these systems. In Sec. III, we propose a network growth model to study the evolving popularity of objects. Then we present the features of user behaviours by calculating the information entropy in Sec. IV. Finally, we discuss our results and the direction of future study in Sec.V.

\section{Statistical properties of Amazon book ratings and Stack Overflow favourite posts}
\begin{figure}[h]
\center\scalebox{0.65}[0.65]{\includegraphics{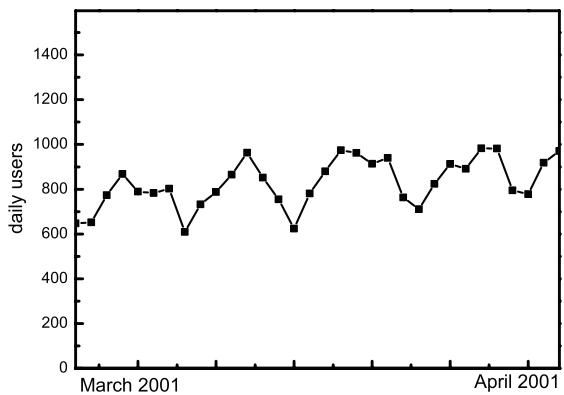}}
\caption{(Color online)Daily number of users of Amazon from March 2001 to April 2001. Similar pattern can be observed in the whole network of Amazon.}\label{regular}
\end{figure}

\begin{table*}
\caption{Basic statistical properties of Amazon and Stack Overflow, including the number of users $u$, the number of the objects $v$, time range and time span.}\label{table1}
\begin{center}
\begin{tabular} {l c c c c c}
  \hline \hline
    Network        &users  &objects    &Rating records       &Time range    &Time span(day)   \\ \hline
    Amazon          &99622      &645055         &2005409       &31st.May,1996-15th.Sep,2005  &3394     \\
    Stack Overflow  &545195     &96680         &1301942                  &2002-2005         &1154  \\

\hline \hline
\end{tabular}
\end{center}
\end{table*}
We analyze two data sets, Amazon and Stack Overflow, to study the statistical properties of evolving bipartite networks, such as the growth of networks and the degree distribution of users. The Amazon is an online shopping service which encourages its users to rate the commodities they buy. The data set of Amazon we analysed is the Amazon book ratings, including 2005409 ratings delivered by 99622 users on 645055 books, from 31st.May,1996 to 15th.Sep,2005. The Stack Overflow is a question and answer website of the Stack Exchange Network, which invites its users to mark their favourite posts. The data set of Stack Overflow has 1301942 ratings, given by 545196 users on 96680 posts, from 2002 to 2005. Table~\ref{table1} gives the basic statistical properties of Amazon and Stack Overflow.
Every entry of these two data sets records the user ID and the object ID with time stamp. We count up the number of ratings in Amazon for each day from March 2001 to April 2001 to study temporal dynamics in the data. The result shows a weekly pattern for the number of ratings of Amazon, shown as Figure~\ref{regular}, which presents the dynamical pattern of human activities.

\subsection{The growth of networks}
The Amazon and Stack Overflow are studied as evolving bipartite networks with two types of nodes,users $U$ and objects $V$. The edges between users $U$ and objects $V$  present the ratings the users giving to the objects.\cite{Barabasi1999} We focus on whether the rating events exist regardless the value of ratings. The large size data with time stamp for long time offer us the chance to reveal the growth of networks, from which we can get an overall view of the development for Amazon and Stack Overflow.

Figure~\ref{Growth} shows the total number of ratings every day in the data sets of Amazon and Stack Overflow respectively, which grow exponentially with time.
\begin{figure}[h]
\center\scalebox{0.65}[0.65]{\includegraphics{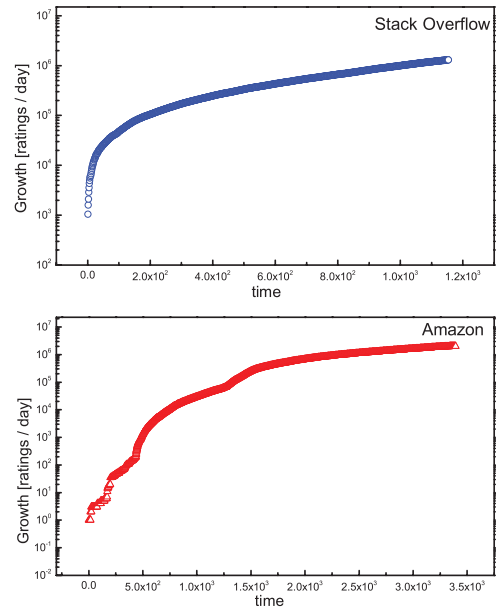}}
\caption{(Color online)The total number of ratings in the data sets of Amazon and Stack Overflow respectively, which increases exponentially with time.}\label{Growth}
\end{figure}

\subsection{Degree distribution}
The degree distribution of users for Amazon and Stack Overflow follow a power law with an exponential cutoff. The degree of the user in the bipartite network is the total number of ratings that the user gives to objects and the degree distribution is the probability distribution of user degrees over the whole network.\cite{degree} To study the aggregated structure, we examine the cumulative degree distributions of users,the result of which fits a power law with exponential cutoff by the method modified by Clauset $et~al$,\cite{modified}
\begin{equation}\label{powerlaw}
\begin{array}{rcl}
F(x)\sim x^{-\alpha}e^{-\lambda x}
\end{array}
\end{equation}
Figure~\ref{Growth} shows the cumulative degree distributions of users for the evolving networks Amazon and Stack Overflow. Table~\ref{table2} shows the value of parameters $\alpha$ and $\lambda$ according to the Eq.(\ref{powerlaw}).
\begin{figure}[h]
\center\scalebox{0.75}[0.85]{\includegraphics{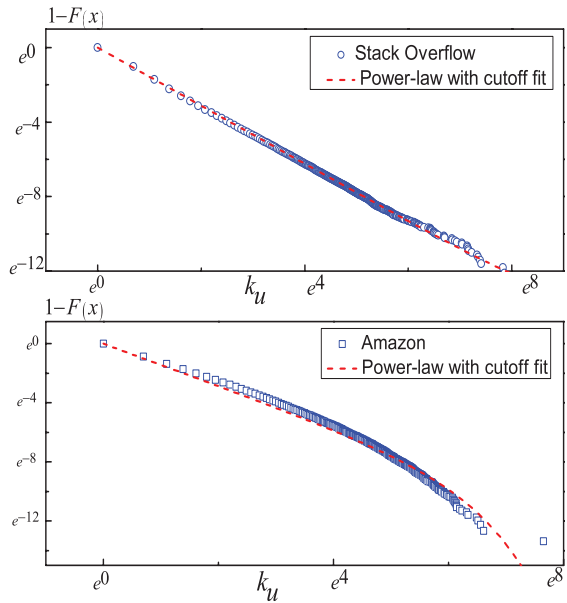}}
\caption{(Color online)The cumulative degree distributions of users for the evolving networks Amazon and Stack Overflow.For Amazon,the dashed curves fitting well to the cumulative degree distributions of users are  with parameters $\alpha=1.4274$ , $\lambda=0.00327$. For Stack Overflow, the dashed fitting curves are with parameters $\alpha=1.5593$ , $\lambda=-0.00112$.}\label{Growth}
\end{figure}

Systems in maturity period are systems which have stable interactions within themselves or with their environments. From the result we can find that, for systems in maturity period, most users are small-degree users while few have large degrees. Users seek gain when rating or finding favourite posts with limited time or energy. In economics, this kind of gain is called the utility and the marginal utility is the gain from an increase, or the loss from a decrease.\cite{utility} When the utility of users achieves equality (marginal utility equals the marginal cost),\cite{marginal} users gradually cease the activities of ratings or marking the favourite posts due to the law of diminishing marginal utility.\cite{diminishing}
\begin{table}
\caption{Fitting parameters of user degree distribution for Amazon and Stack Overflow.}\label{table2}
\begin{center}
\begin{tabular} {l c c c c}
  \hline \hline
    Network        &mean $\alpha$   &variance $\alpha$    &$\lambda$     &variance $\lambda$  \\
    \hline

    Amazon          &1.4274     &0.0077                  &0.0033                &0.00153  \\
    Stack Overflow  &1.5593     &0.0012                 &-0.00112               &0.00012 \\

\hline \hline
\end{tabular}
\end{center}
\end{table}

\section{The evolving popularity of objects}
The popularity of objects is decided by complex reasons and evolves with the network growth. Evolving networks change with time, not only by adding or removing links, but also by adding or removing nodes, which are closest to natural networks. The first evolving network model was BA model proposed by Barab¨¢si Albert to study the scale free networks,\cite{Barabasi1999} including two significant concepts, growth and preferential attachment. Based on the BA model, we use a model to study the evolving popularity of objects. Regarding to the system science, systems have varied states with different structures. In this paper, we focus on systems in maturity period where the evolution of popularity for objects can be more representative.
\subsection{The model for popularity of objects}
In this section, we propose a dynamical model to investigate the evolving popularity of objects. According to the well-known model of network growth, Preference Attachment, the probability of a node $v_{i}$ with degree $k_{i}$ acquiring new contacts is
\begin{equation}\label{PA}
\begin{array}{rcl}
P(v_{i})=k_{i}/\sum_{j=1}^Nk_{j}
\end{array}
\end{equation}

In our model, we use Relevance $R_{i}$ to show a regular way how popularity of objects evolve, where the degree of object $i$ at time $t$ is defined as $k_{i}(t)$. Relevance $R_{i}$ is the ratio between the real number of ratings one object received and the expected number of ratings the object will receive. We assume that $X(t,\Delta t)$ is the number of new ratings added to objects in the network during the next $\Delta t$ days. What need to be noticed is that the total number of objects varied with time. So $m$ is a variable with time $t$, used to indicate the number of existed objects. The expected number of ratings that object $i$ will receive during next $\Delta t$ can be denoted as
\begin{equation}\label{expected}
\begin{array}{rcl}
 \Delta k_{i}(t,\Delta t)=X(t,\Delta t)k_{i}(t)/\sum_{j=1}^{m}k_{j}(t)
\end{array}
\end{equation}
 And we use $\Delta K_{i}(t,\Delta t)$ to represent the real number of ratings that object $i$ received. Thus, the Relevance $R_{i}$ is shown as Eq.(\ref{ratio}).
\begin{equation}\label{ratio}
\begin{array}{rcl}
 R_{i}(t,\Delta t)=\Delta K_{i}(t,\Delta t)/ \frac{X(t,\Delta t)k_{i}(t)}{\sum_{j=1}^{m}k_{j}(t)}
\end{array}
\end{equation}

However, we ignore the initial time of networks when each object receive its first rating. The development of networks at initial time is complex which is related to the "cold boot". To better clearly display how the popularity of objects evolves, we assume that the $k_{i}(t)\geq1$ and no time periods have no ratings,that is, $C(t,\Delta t)\neq0$.

\begin{figure}[h]
\center\scalebox{0.65}[0.65]{\includegraphics{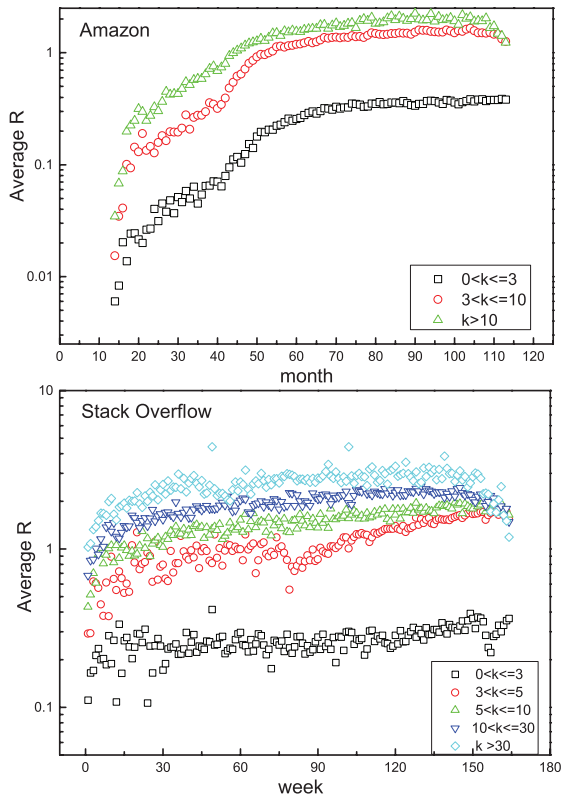}}
\caption{(Color online) The evolution of objects' popularity for Amazon and Stack Overflow. For Amazon, the objects are divided into three groups by their degrees with $\Delta t=30 days$. When the system goes into a maturity period, the average Relevance $R(i)$ of the group $k>10$ fluctuates around 2 while the Relevance $R$ of the group $0<k\leq3$ is about 0.36. For Stack Overflow, the objects are divided into five groups based on their degree with $\Delta t=7 days$. The value of the Relevance $R$ of the group $k>30$ is about 3 as the group $0<k\leq3$ only fluctuates around 0.27.}\label{popular}
\end{figure}

\subsection{The empirical study of the evolving popularity of objects for Amazon and Stack Overflow}
We apply the empirical data of Amazon and Stack Overflow to our model. For Amazon data, we first divide all objects into three groups depended on their degrees as $0<k\leq3$, $3<k\leq10$, and $k>10$. Then, we separate the data into 114 time windows to see how it evolves with time. For Stack Overflow data, we group the objects into five types based on their degrees as $0<k\leq3$, $3<k\leq5$, $5<k\leq10$, $10<k\leq30$, and $k>30$ and divide the sequence of time into 165 time windows. We calculate the average Relevance $R_{i}$ of different groups for Amazon and Stack Overflow, respectively. It can be clearly seen that the average Relevance $R_{i}$ increases exponentially at the initial time. Then, $R_{i}$ starts to increase slowly until the networks evolve to a certain stage, the maturity period. It is the time when networks have stable interactions within themselves or with their environments. So $R_{i}$ will fluctuate around a certain value in that period. However, the $R_{i}$ of objects with large degree will decrease after a certain time, since the number of ratings received by objects with large degree arrives a saturate level.

From the above results, we find that not all the evolution of popularity for objects fit the PA method and only objects with not too large or small degree fit the PA method well. The number of ratings given to objects in next period is not completely driven by their current popularity, which is influenced by other exterior reasons, like the environment of the market and limitation of the user's number. The popularity of objects in each group exhibits heterogeneous fitness values which increase with the growth of evolving bipartite networks, shown as Figure~\ref{popular}. The larger the degree of one object is, the more ratings it will receive compared to the expected number of ratings calculated by Eq.(\ref{expected}). Finally, due to the limitation of the user's number, the number of ratings received by large-degree objects would first achieve the saturate level so that the objects' popularity will gradually decay. Figure~\ref{popular} shows an evolving process of the objects' popularity.

\section{The evolving diversity of user behaviours}
In this section, we discuss the features of user behaviours by calculating the information entropy. To operate and manage an open system, one of the key points is to understand user behaviours. We need to predict the user behaviours by collecting and analyzing the potential signals so that we can ensure the systems are under control. Moreover, for e-business systems, learning about the user behaviours can help them to upgrade the experience of the users,\cite{T,Liu,Liu2} make personal recommendation and provide customized advice to users.
\subsection{The information entropy}

Entropy is introduced to measure disorder or uncertainty. Shannon entropy was first proposed by Claude Shannon in 1948 to measure the unpredictability of information content.\cite{shannon} Warren Weaver extended the meaning of the "information" mentioned by Shannon. He pointed out that the word information in communication theory is not related to what you do say, but to what you could say. That is, information is a measure of the diversity of one's choice.\cite{shannon2} For instance, a choice $H$ can lead to $n$ results while the possibility of each result is expressed as $p_{1}, p_{2},\ldots, p_{n}$. The information entropy of this choice can be shown as Eq.(\ref{entropy})
\begin{equation}\label{entropy}
\begin{array}{rcl}
H=H(p_{1},p_{2},\ldots,p_{n})=\sum^{n}_{i=1}p_{i}\log_{2}\frac{1}{p_{i}}
\end{array}
\end{equation}
As we known, the larger the information entropy is, the more uncertain the system is. And the information entropy can be used to measure the diversity of user behaviours.\cite{zhang} We group the objects and users by their degrees that each group of users and objects has same degree value. Then, we measure the diversity of users in each group to display the features of user behaviours for the whole system. The information entropy of users with degree $k$ is shown as Eq.(\ref{entropy2}). We assume $k$ is the degree of user; $n(k)$ is the total number of objects with different degrees, chosen by users with degree $k$; $p_{i}$ is the probability that objects with degree $i$ chosen in total ratings times.
\begin{equation}\label{entropy2}
\begin{array}{rcl}
H(k)= \sum^{n(k)}_{j=1}p_{i}\log_{2}\frac{1}{p_{i}}
\end{array}
\end{equation}
$p_{i}$ can be calculated according to the PA model, as shown in Figure~\ref{PA2}. That is, the possibility of objects in each group can be calculated depending on each group's degree. $K_{i}$ is the total degree of objects with degree $i$ and $N$ is the count of the users with different degrees.
\begin{equation}\label{PA2}
\begin{array}{rcl}
p_{i}=K_{i}/\sum_{j=1}^NK_{j}
\end{array}
\end{equation}
Thus, the larger the information entropy is, the more diverse the user behaviours are. It is appropriate to display the features of user behaviours in bipartite networks by the information entropy.
\subsection{The empirical study of user behaviours for Amazon and Stack Overflow}
We do empirical study of the user behaviours in the data sets Amazon and Stack Overflow by calculating the information entropy. First, we group the users and objects by their degree for Amazon and Stack Overflow respectively. Then we employ each data sets to Eq.(\ref{entropy2}). To better display the features of user behaviours, we normalized the information entropy by the principle of maximum entropy. As we known, it is a natural property for systems to develop toward to disorder. So the choice with largest entropy can best represent the state of other choices \cite{J1,J2} and it is the time when $p_{i}$ is same. The $\theta(k)$ is the normalized information entropy, shown as Eq.(\ref{maximum}).
\begin{equation}\label{maximum}
\begin{array}{rcl}
\theta(k)= \sum^{n(k)}_{j=1}k_{i}\log_{2}\frac{1}{p_{i}}/\log_{2}n(k)
\end{array}
\end{equation}

\begin{figure}[h]
\center\scalebox{0.75}[0.75]{\includegraphics{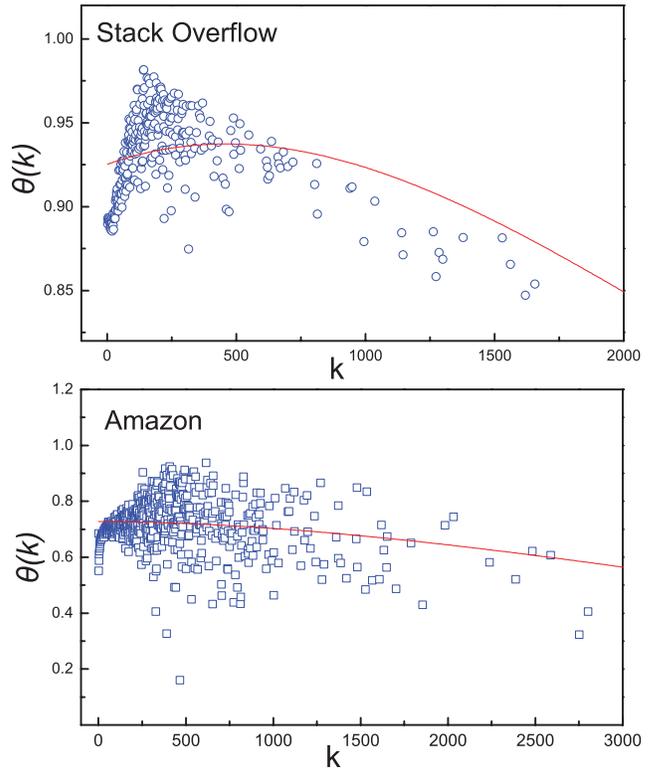}}
\caption{(Color online) The normalized information entropy $\theta(k)$ of the user with different degree $k$ for Stack Overflow and Amazon.}\label{information}
\end{figure}

Figure~\ref{information} shows that $\theta(k)$ abruptly increases as the user degree increases initially, which can be up to nearly 0.92 and 0.98 for Amazon and Stack Overflow respectively, when 1 is the time that no preference exists in user behaviours at all. Then the information entropy decreases gradually as the degree of users increases. The result illustrates a phenomenon that users tend to choose different types of objects when they first enter the network. Then gradually their behaviours are acted with more and more strong purpose, which means that users start to know clearly about what they want. We fit the results by the least square method. Thus, there is an apparent trend that shows the decreasing inclination of the normalized information entropy, that is, the interests of user behaviours tend to be centralized as their degrees raise. Generally speaking, the interests of large-degree users are more centralized than small-degree users.

\section{Conclusion and discussions}
\subsection{Summary}
In this paper, we investigate the evolving properties of systems by empirical study of the Amazon and Stack Overflow. To better display the different aspects of the systems, we abstract the Amazon and Stack Overflow as bipartite networks, studying the development of these systems as well as the weekly pattern of rating behaviours and the degree distribution. We also find that the user degree distributions of Amazon and Stack Overflow follow a power law distribution with an exponential cutoff.

Moreover, we presented a network growth model, in which we can learn about the evolving popularity of objects. Based on the idea of Preference Attachment, the average relevance $R(i)$ is applied to reveal the evolvement of rating process. By empirical analysis, we find that the popularity of objects in each group exhibits heterogeneous fitness values which increases with the growth of evolving bipartite networks. The objects with large degree will receive more ratings compared to the expected number of ratings based on PA method and vice versa. The objects with not too large or small degree can best fit the PA method. Current popularity cannot be completely accounted as the reason deciding the number of ratings that objects receive. Furthermore, we study the properties of evolving networks from the perspective of users. The information entropy is applied to the data sets of Amazon and Stack Overflow. The users in these two networks present a common phenomenon, that the larger degree the user have, the stronger purpose are with their behaviours. In conclusion, this paper discusses the properties of networks from the aspects of users, objects and rating behaviours, which gives advice to the system operators on how to face the challenges brought by the evolution of bipartite networks.

\subsection{Limitations and future work}
Although our paper discusses different aspects of the evolving bipartite networks, it also has a few defects:
Firstly, in this work, the network growth model can only display a simple process how the popularity of objects evolve, lacking details analyzing the specific factors that influence the evolvement process. Regardless of the complexity and diversity of the user behaviours, this model cannot be a sufficient basis to explore thoroughly the evolving popularity of objects in bipartite networks. Therefore, the model of network growth needs further research in the future work.

Secondly, from the result of our network growth model, we can see that the relevance $R(i)$ of the groups with largest degree for Amazon and Stack Overflow both decrease after a certain time. As another direction of the future work, we need to explore deeper about when objects with large degree will cease to acquire more ratings. If we can predict the time when the popularity of welcomed objects will decay by monitoring the relative index, it will have remarkable influence on the business research.

Thirdly, it is worth pointing out that we study diversity of user behaviours in the aggregated networks without comparing the difference of the diversity of user behaviours at different time. In this sense, the dynamical user behaviours would be a key issue for our future work.

\begin{acknowledgments}
We thank Kai Yang for useful comments and suggestions. 
\end{acknowledgments}

\end{document}